# Advanced Customer Activity Prediction based on Deep Hierarchic Encoder-Decoders

Andrei Ionut Damian
*Lummetry.AI*
Bucharest, Romania
andrei@lummetry.ai

Nicolae Tapus
*University Politehnica of Bucharest*
Bucharest, Romania
ntapus@cs.pub.ro

Laurentiu Piciu
*Lummetry.AI*
Bucharest, Romania
laurentiu@lummetry.ai

Sergiu Turlea
*Lummetry.AI*
Bucharest, Romania
sergiu@lummetry.ai

*Abstract* — Product recommender systems and customer profiling techniques have always been a priority in online retail. Recent machine learning research advances and also wide availability of massive parallel numerical computing has enabled various approaches and directions of recommender systems advancement. Worth to mention is the fact that in past years multiple traditional "offline" retail business are gearing more and more towards employing inferential and even predictive analytics both to stock-related problems such as predictive replenishment but also to enrich customer interaction experience. One of the most important areas of recommender systems research and development is that of Deep Learning based models which employ representational learning to model consumer behavioral patterns. Current state of the art in Deep Learning based recommender systems uses multiple approaches ranging from already classical methods such as the ones based on learning product representation vector, to recurrent analysis of customer transactional time-series and up to generative models based on adversarial training. Each of these methods has multiple advantages and inherent weaknesses such as inability of understanding the actual user-journey, ability to propose only single product recommendation or top-k product recommendations without prediction of actual next-best-offer. In our work we will present a new and innovative architectural approach of applying state-of-the-art hierarchical multi-module encoder-decoder architecture in order to solve several of current state-of-the-art recommender systems issues. Our approach will also produce by-products such as product need-based segmentation and customer behavioral segmentation – all in an end-to-end trainable approach. Finally, we will present a couple methods that solve known retail & distribution pain-points based on the proposed architecture.

## I. Introduction

According to various published research such as [1] [2] it is well known that the online medium has served as a powerful driving force for the development of recommender systems technologies due to the exponential adoption of online business transactions. Market research sources, such as *Markets & Markets Inc.* and *Reportsnreports.com*, state that recommendation engines market including, but not limited to, well known methods such as collaborative filtering, content-based filtering, hybrid recommender systems is forecast to reach $4414.8M by 2022 from a $801.1M in 2017 at a CAGR of 40.7%. It is obvious that this increase is mainly driven by increase in focus toward enhancing consumer experience not only in online environment but also in traditional offline business such as classic retail stores. Beside the market hunger for more advanced consumer experience based on behavioral analytics another driver is due to the technological advancements in machine learning in general and in deep learning in particular.

Our work is targeting the general areas of hybrid recommender systems and business predictive behavioral analytics. In this paper we will argue that our proposed models solve several known issues related to in-session and multi-session collaborative filtering systems with particular focus of capturing the consumer behavior both at session level and throughout the whole customer-lifetime.

Due to the fact that matrix factorization methods or *word2vec* [3] based methods such as [4] fail to capture the consumer-journey and are more focused on static user behavioral representation, in past years sequence-oriented recommender systems based on deep recurrent directed acyclic graphs (DAG) has seen increased attention with multiple approaches such as [5] [6] [7] [8]. Also, another important research area that strongly relates to our work is that of neural language models and sequence-to-sequence [9] in particular, as it will be shown in the following sections.

The proposed architecture combines several approaches in order to create a model capable to both understanding in-session inter-dependencies and also session-to-session context-oriented dependencies while modelling two different latent spaces - that of products/services and the user base. While some current state-of-the-art research such as [10] [11] focuses on determining efficient hits or single next-hit [12] within the overall recommendation basket, our goal is to capture intra-session basket patterns as well as user lifetime behavioral patterns and seasonal/periodic recurrent ones.

This overall approach enabled us during our experiments to generate multiple reliable results for various tasks such as time-to-next-event prediction and sequence decoding of the next customer session basket content. Nonetheless, more simple tasks were experimented, such as understanding products (or services) needs-oriented clustering and generate behavioral segmentation of customers on top of the learned latent-space vector embeddings. Another area of successful experimentation area has been that of behavior pattern anomaly detection that will be further presented in *Section IV.B*. All of this has been achieved with a single end-to-end trainable hierarchical sequence-to-sequence DAG as it will be further presented in the Approach section.

Finally, the proposed architecture manages to extract meaningful and interpretable results from a user journey funnel transactional data as we will see in the later sections.

## II. RELATED WORK

As previously mentioned, our work strongly relates to two different areas of research and development – that of sequence-based recommender systems and also to seq2seq [9] neural language models with particular focus on hierarchical recurrent models [13].

Motivated by the fact that traditional approaches such as collaborative filtering and matrix factorization methods consider the user as a static entity with fixed interests through time, many researchers started to frame the recommender systems as sequence-based problems. For example, Devooght et al. [8] proposed a vanilla many-to-one architecture based on Gated Recurrent Units (GRUs) [14] which process each market basket as a products timeseries, where each product is encoded as an one-hot vector. The output is a dense layer with the number of neurons equal to the number of products which computes a softmax function. Finally, the most likely *k* products to be of interest to a user are the k items whose neurons are activated the most.

In a similar manner as the one presented above, Hidasi et al. [6] used a stacked-GRU network and experimented also using one-hot encoded product vectors and random initialized trainable embeddings in order to address the sparsity of the input.

Seq2seq gained popularity since 2014, when Sutskever et al. [9] proposed an innovative architecture for general sequence learning. Their approach managed to resolve a limitation of Deep Neural Networks which could only manage problems that output a fixed-length vector. However, there were many important problems that could be better represented using variable-length sequences (NMT, question answering etc.). Therefore, they proposed an end-to-end architecture composed of an **encoder** (which encodes the timeseries input into a fixed-length vector) and a **decoder** (which maps the encoding to the target timeseries). It is known that a DNN that can process sequences (Recurrent Neural Networks [15] [16]) can easily map sequences to sequences whenever the alignment between the inputs and the outputs is known ahead of time. Finally, decoupling the architecture into two separate RNNs leads to a more general strategy which allows to map input sequences to output sequences, independent of their sizes. Based on this aspect, in our work we will be able to predict the customer next session full content of products/services by understanding (encoding) the customer's lifetime behavior.

Sordoni et al. [13] proposed a more general encoder architecture which is stacked into two parts:

1. A "child" RNN which discovers the in-session inter-dependencies;
2. A "parent" RNN which processes all the in-sessions inter-dependencies and maps to a fixed-length vector the session-to-session context aware inter-dependencies.

## III. APPROACH

This current approach is built starting from our previous works [17] [18]. The first one [17] presents an end-to-end model that is capable to process sequential events (user-interactions, transactions etc.) in order to find real-valued representations of each product/service and each user, such as they lie in latent vector spaces (products and users latent spaces). The second one [18] presents two different models used for churn prediction, with the second model (time-to-next-event prediction) being improved in this work (III.B).

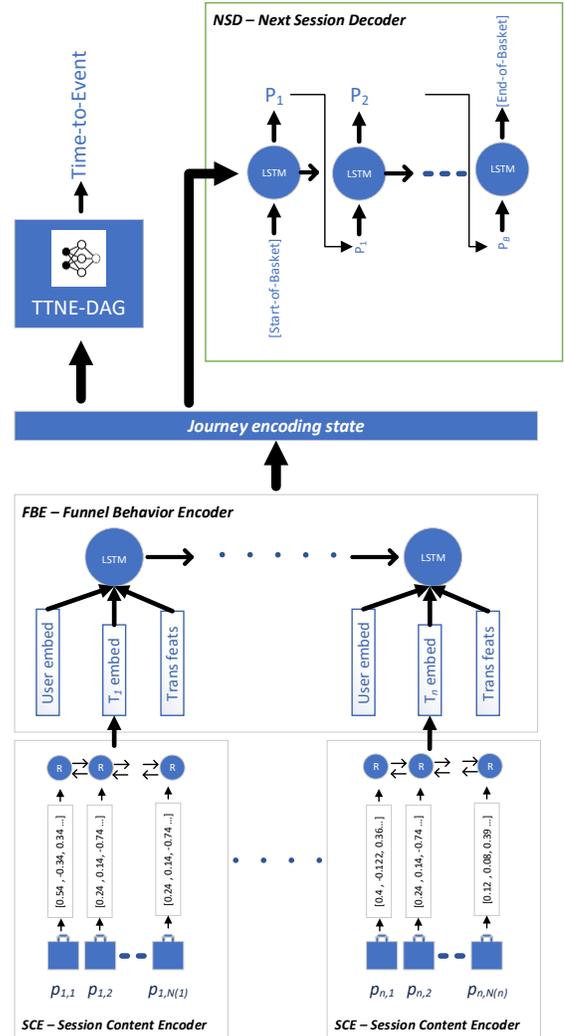

*Figure 1 The user journey funnel Deep Hierarchic Encoder-Decoder architecture where each session basket $T_i$ with i:[0..n] has N(i) products that are encoded by the SCE RNN and passed together with session-level features to the FBE RNN encoder. The final output of the encoder is the either decoded by the NSD predictor or passed to the time-to-next-event FC regressor*

*A. Products and users latent spaces*

Our end-to-end trainable DAG presented in Figure 1, which is particularly designed as a recommender system, uses in this case as main inputs the multi-session time-series of products/services that are bought by a customer. All the discrete inputs such as item codes, user ids, dates, are one-hot encoded and by the input layers of the DAG. Nevertheless, in order to enrich the model capability to understand and discover behavioral patterns we need to inject additional information about these inputs. As a result, we are using both item-embeddings as well as user-embeddings. For both item-embeddings and user-embeddings we are using two different scenarios as presented below:

- "*cold-start*": where each latent space item-embedding is initialized randomly and re-positioned within the vector spaces just by using the task optimization function. For the particular case of user encodings, we are using only cold-started embeddings sampled randomly uniform;
- "*warm-start*": is the second case where each latent space vector was previously trained (using the DAG presented in our previous work [17]) in order to make the *item semantic information* more accessible from the very beginning of the optimization process to the end-to-end model. In our tests, this setting revealed better results, in particular for the product/item embeddings, results that will be presented in the next section. This "*warm-start*" scenario has also the option of allowing or not the further fine-tuning of the item-embeddings during the end-to-end model optimization process by allowing or blocking the gradient propagation within the embeddings matrix.

*B. Hierarchical time-to-next-event prediction*

In order to further optimize our previously proposed architectures for the DAG used for time-to-next-event prediction (as presented in our [18]), we propose a similar approach to that of Sordoni et al. [13]. Thereby, in order to capture more semantical information about each market basket that is processed at each timestep, we dropped the old encoding methodology of the products/services (averaging the embeddings at each timestep) and replaced it with a Session Content Encoder (*SCE*) model based in our current experiments on RNN cells (in particular, LSTM [19]). The *SCE* module processes bidirectionally each market basket by on-hot encoding each item and then looking-up its embedding within an item-embeddings matrix. This provides the Funnel Behavior Encoder (*FBE*) – also based on RNN cells in our experiments - a better representation of the products/services than just a simple average/sum. Thus, the *FBE* is able to understand much better the inter-session dependency for returning customers and thus create an actual representation of the user journey. In turn, the *FBE* module receives additional session-level information such as date-related features as well as funnel-level information such as user ids. As previously mentioned, in our architecture we use latent space embeddings for each user with the specific purpose of capturing the user-journey information in each individual user-embedding. Each user-embedding is initialized randomly uniform and jointly optimized during the training with the rest of the DAG weights. As mentioned in previous sub-section, the optimization process includes also the item-embeddings in some of our scenarios developed in order to determine the pre-trained & refined item-embeddings efficiency vs pre-trained only embeddings setup.

Finally, the actual prediction is generated by a small fully connected network that has a simple regression unit at the output layer which generates the time until next event prediction. This allows us to use a simple distance-based loss function such as mean squared error or mean absolute error

*C. Hierarchical encoder-decoder*

Motivated by the state-of-the-art results obtained by seq2seq methods in addressing problems such as NMT or our own neural generative language models (chatbots), we propose an innovative architecture for recommender systems based on hierarchical encoders-decoders. Therefore, using the proposed hierarchical modules *SCE – FBE* presented in III.B, we coupled a decoder - Next Session Decoder (*NSD*) - module that aims to predict the next customer session full content, without any restrictions of content dimensionality, all this due to its auto-regressive nature.

*D. Optimization process*

The training process follows the classical approach of maximizing the probability, represented by the log-likelihood, of decoding the correct session content using teacher-forcing on the *NSD* module and generating continuously incrementing hierarchic time-series for each individual user with a minimal threshold of transactions per funnel. More precisely we have the objective of finding the *θ* model parameters that satisfy the below equation where $Y_{NSD}, X_{SCE}, X_{FBE}$ are all time-series slices from the training dataset *D*. Our final result is that of jointly optimizing all three proposed neural modules based on the hierarchically encoded funnel from first session up to the last session and the next session target content that is both fed as end-to-end model target as well as NSD input.

$$\underset{\theta_{SCE,FBE,NSD}}{\operatorname{argmax}} \sum_{(Y_{NSD}, X_{SCE}, X_{FBE}) \in D} \log p(\hat{Y}_{NSD} | Y_{NSD}, X_{SCE}, X_{FBE})$$

IV. EXPERIMENTS AND RESULTS

*A. Our experimental approach*

*The proposed architectures*

For the current presented experiments we chosen multiple different architectures, ranging from the simplest one up to the most resource demanding one. In this paper we will analyze the results of simplest model (LENS1000) and a more complex one (LENS2000). The basic core architecture details can be summarized in the below *Table 1*

| Model | SCE layers | SCE cells sizes | FBE layers | FBE cells sizes | NSD layers | NSD cells sizes |
|---|---|---|---|---|---|---|
| LENS1000 | 1(b) | 64 | 1(b) | 256 | 1 | 512 |
| LENS2000 | 1(b) | 256 | 2(b) | 256 | 2 | 512 128 |

*Table 1 - Architecture details for each of the three modules of the benchmarked models. Note that bidirectional recurrent layers are noted with (b) and the cell size applies to all module layers if otherwise noted*

*The datasets*

In order to train, validate and finally benchmark our models we have used the following public available market baskets datasets, namely *ta-feng* and *foodmarket,* together with real-life datasets from our customers.

*Ta-feng* dataset contains 817,741 transactions belonging to 32,266 users which jointly bought 23,812 unique items. On the other hand, *foodmarket* dataset is a sample dataset and it contains a limited number of transactions (54,537) belonging to 8,736 users and just 1,560 unique products. Both datasets have the following minimal structure: *TRAN_ID, CLIENT_ID, PROD_ID, TIMESTAMP, PROD_AMOUNT, PRODT_QTY*. However, due to the limited amount of funnel information that the *foodmarket* dataset provides we decided against relaying on the its resulted performance indicators. As a result, in the current paper only results from the *ta-feng* dataset are available.

In order to feed our proposed Deep Hierarchic Encoder-Decoder architecture, we grouped the data in transactions timeseries which describe the buying behavior for each individual customer (i.e. each timestep $i$ specifies which are the products bought by a particular customer in their transaction $i$, as well as other features such as products amounts and quantities, time from previous purchase, timestamp encoding).

*The training methodology*

To summarize the details of the training methodology we have to mention two different aspects: the pre-training approach for the SCE input embeddings (which is described in [17]) and the training cycles of the end-to-end models.

The end-to-end models were constructed and trained using Tensorflow [20] with GPU capabilities, which allows us to deploy the LENS+ backend API (including computational graphs) in production-grade systems where it provides day-to-day predictions and continuous training. The parameters of the models were optimized using RMSprop Gradient Optimization with a learning step equal to 0.001. In order to use the GPU memory and the offload procedure more efficiently the training observations were grouped in batches of maximum 128 samples.

### B. Behaviour anomaly detection

One of the main targets of our experiments has been that of generating automated smart-insights based on the direct and indirect results of our architecture. The main experiment has been that of detecting anomalous signals – such as a customer or a meta-customer that has a change in behavior patterns. For this purpose, we designed a simple process/algorithm defined below

---
**Algorithm** 1 *AnomalyDetector*($M$, $D$, $f_k$, $d_K$, $d_A$, $d_B$, $f_E$)
---
for each funnel $U_i$ ($i:[0..N]$) time-series in $D$:
  predict $P(U_{i,T+1}|\ U_{i,0:T},\ M) \Leftarrow$ next element in funnel $C_i$
  observe $(U_{i,T+1}) \Leftarrow$ next element in funnel $C_i$
  compute $d_A(P_i,R_i) = |P(U_{i,T+1}),\ R(U_{i,T+1})|$ normed distance
end for
$C \Leftarrow f_k(f_E(i\ |\ M)\ |\ D,\ d_K\ |\ i:[0..N])$
$O \Leftarrow \{0\ for\ all\ i:[0..N]\}$
for each cluster $Cj$ with $j:[0..K]$
  for each funnel $i$ in all $C_j$ funnels:
    $Oi \Leftarrow d_B(d_A(P_i,R_i)\ ,\ d_A(P_j,R_j)\ |\ j \neq i,\ j:[0..N_j])$
**Return** $O$
---

Basically, our algorithm uses a full pipeline $M$ that actually consists encodes each session with the *SCE* module, prepares the full funnel encoding with the *FBE* module and finally generates the content with the full *NSD* module or just a regression head. Other elements are the $D$ data-series, a clustering function $f_k$, three distance functions $d_K$, $d_A$ and $d_B$ and finally a function $f_E$ that generates funnel embedding by extracting this information from $M$. The first stage of the algorithm consists in determining the *distance* between prediction and actual observed information with a distance function $d_A$ that also applies a normalization approach to all inferred distances. The second stage will employ the clustering function $f_k$ that uses the embedding extraction function $f_E$ in order to compute the optimal clustering $C$ with $K$ clusters for all $N$ funnels. The third and final stage will compute a pairwise distance using the distance-like function $d_B$ between all funnel distances computed at stage one with the distance function $d_A$. This will be executed within each individual cluster in order to detect the potential funnels where the variation between the predicted step and the real observed step is identified as outlier in comparison with the similar behaving funnels within the cluster.

### C. Out-of-stock (OOS) preventive alerts

Another practical experiment of our *SCE-FBE-NSD*-based system if that of generating out-of-stock signals. We designed an algorithm which is presented below (**Algorithm** 2) that uses both the direct results of our architecture (next basket prediction and time-to-next-event prediction for each customer) and the a priori known resupply matrix for a certain number of days. The actual application is used by retailers, merchandisers and distributors who receive from the system out-of-stock alerts based on the residual matrix (the difference, for each product and each day, between the resupplies and predicted stocks).

**Algorithm** 2 *OOSAlerts(M, D, resupplies[n_prods][n_days])*

$pred\_stocks \Leftarrow zeros[n\_prods][n\_days]$
for each funnel $U_i$ (i:[0..N]) time-series in D:
  predict $P(U_{i,T+1}/ U_{i,0:T}, M), Q(U_{i,T+1}/ U_{i,0:T}, M) \Leftarrow$ next basket of $U_i$ //products and quantities
  predict $T(U_{i,T+1}) \Leftarrow$ time-to-next-event for $U_i$
  if $T(U_{i,T+1}) < n\_days$:
    for each product, qty $p_j, q_j$ [j ≥ 0] in $P(U_{i,T+1}), Q(U_{i,T+1})$:
      $pred\_stocks[p_j][ T(U_{i,T+1})] \Leftarrow$
                      $pred\_stocks[p_j][T(U_{i,T+1})] + q_j$
    end for
  end if
end for
$R \Leftarrow resupplies - pred\_stocks$
$A \Leftarrow \{(i,j)$ for which $R[i][j] < 0\}$
**Return** $A$

Although our entire presented approach is focused on customer-journey behavior analysis, for this particular case of out-of-stock prediction we have devised a strategy for the customer-agnostic scenarios. The algorithm for the customer-agnostic scenario is based on partially reframing the problem and the particular hierarchical structure of the customer-oriented data-funnels as location/store transaction funnels.

*D. Results*

Finally, the actual results of our experiments can be summarized in the benchmarking *Table 2*. Due to our experiment objective of predicting the next basket content we employed the *Recall*, *Precision* and *F1* scores. The *Recall* score basically captures the basket coverage of our predictions, the *Precision* determines the efficiency of our predictions and finally the *F1* score generates the aggregated score. We opted not to use HR@k (Hit@k) as this particular metric assumes a series of fixed k items in real series and in the predicted sequence, however our model has the purpose to predict the variable size basket at each generative auto-regressive step.

$$Recall = \frac{\sum_{p_i \in P_B} m(p_i, B)}{|B|}$$

$$Precision = \frac{\sum_{p_i \in P_B} m(p_i, B)}{|P_B|}$$

In the above formulas *m* is a function that returns 1 if the predicted $p_i$ product from the full predicted basked $P_B$ is found in the actual *B* basket.

All the results on our models are achieved on a validation set representing the last baskets of randomly selected 30% of the customers in the original dataset. These baskets were not provided to the model during the training process.

| Dataset | Ta-Feng dataset | | |
|---|---|---|---|
| Models | Recall | Precision | F1 |
| XGBoostClassifier baseline | 0.0948 | 0.2779 | 0.1414 |
| ATEM [21] | 0.1089 | - | - |
| LSDM [12] | - | 0.1237 | - |
| ANAM [22] | - | - | 0.1460 |
| **LENS1000** | **0.1138** | **0.3723** | **0.1743** |
| **LENS2000** | **0.1255** | **0.4032** | **0.1914** |

*Table 2 Result of our experiments with the two architectures on the Ta-Feng dataset.*

For the above comparison results we used scores computed on multiple recommendation lists (with maximum size *k*=10) and thus the average scores were taken.

If precision was not available in the external results, the hit-ratio (*HR@k*) was considered instead, as within our experiments and for our purpose they actually do represent the same metric.

V. CONCLUSIONS AND FURTHER WORK

*A. Conclusions*

Current state-of-the-art architectures used for recommender systems leverage the capability of RNNs to model temporal dynamic behavior and, therefore, capture the intrinsic properties present in customers' previous purchases and interests. However, they lack in generating accurate abstractions when the setting implies sequential data whose events are comprised of several components (e.g. a basket of purchased items). Moreover, the models employed to this extent are only capable of producing recommendations of single products or top-k most likely products, rather than predicting complete future purchases.

We overcome these limitations by proposing a hierarchical model in which we "combine" two RNNs - the (parent) RNN focusing on the customer's shopping history and child one targeting transactions at the session (basket) level. This approach allowed us to obtain from the "child" bidirectional RNN an actual embedding representation of each transaction content far better from other methods. For this particular purpose the classic approach is to combine the basket items by averaging basket items embeddings that usually destroys important information such as basket size, individual identity information for complex baskets and also least important information such as basket ordering.

At the same time, taking inspiration from seq2seq models we coupled a decoder with the purpose of predicting complete future transactions for a customer, without any restrictions of content dimensionality.

## B. Further ongoing work

Our current research focus is geared both towards improving our results by employing multi-stage training-retraining of our models, adopting and adapting other architectures that have been proven succesfull in NMT as well as enabling our end-to-end model to jointly predict other important insights such as the time-to-next-purchase more efficiently than our previous work *[18]*.

*Jointly prediction of basket content and time-to-purchase*

As described in our previous sections our current proposed architecture uses a deep hierarchical DAG encoder to generate the *current state* of the user funnel at each individual step of the purchasing sessions timeseries. This state is then either propagated through a regression FC module for the prediction of time-to-next-event or passed to a decoder DAG based on RNN cells such as LSTM or GRU in order to apply a simple autoregression mechanism for full session basket decoding. Our current experimentation at the time of this paper publishing is focused on the jointly optimization of both above objectives. The target of this further research and experimentation is to obtain a model that achieves at least similar results with that of twin model approach - where the regression FC module and the *NSD* module (sequence decoder DAG) are optimized separately.

*Reinforcement learning online fine tuning*

The next step in our research and experimentation related to this area advanced user journey prediction is to include two-stage training of the proposed models: the first one being the current supervised method and the second one being a fine-tuning stage using reinforcement learning approach. We have reasons to believe that by swapping the normal optimization process, after a fixed period of training, with one that involves reinforcement learning approaches, we might greatly impact the overall results. In this regard, we plan on using policy gradient methods and building a reward shaping function that would take into consideration the length of the predicted output and the number of correctly predicted items in the transaction.

In terms of general approach, we plan to apply this fine-tuning step not on the whole *SCE-FBE-NSD* structure but rather isolate the *FBE* module from the overall architecture and focus on re-training in the reinforcement setting only this particular module of the overall DAG. As a result, we plan to break-down the *FBE* module within the actual funnel encoder and a policy agent that will use the funnel encoding as the state of the observed environment.

*Employing Transformer architecture*

Even if the results of the hierarchical encoder-decoder are very promising, we believe that an attention mechanism would bring improvement in the developed recommender system. However, we are going to completely change the current architecture, using besides reinforcement learning online fine tuning, the Transformer architecture proposed in 2017 by Vaswani et al. [23] which produced state-of-the-art results in seq2seq tasks. As a result, we are currently experimenting with "transforming" each of the three main modules into their Transformer-architecture counterparts and further testing this new approach.

REFERENCES

[1] C. C. Aggarwal, Recommender systems, Springer International Publishing., 2016.

[2] N. Polatidis and C. Georgiadis, "Recommender Systems: The Importance of Personalization in E-Business Environments," *International Journal of E-Entrepreneurship and Innovation,* no. 0.4018/ijeei.2013100103, pp. 32-46, 2013.

[3] T. Mikolov, I. Sutskever, K. Chen, G. Corrado and J. Dean, "Distributed representations of words and phrases and their compositionality," in *proceedings of the 26th International Conference on Neural Information Processing Systems - Volume 2 (NIPS'13), C. J. C. Burges, L. Bottou, M. Welling, Z. Ghahramani, and K. Q. Weinberger (Eds.), Vol. 2. Curran Associates Inc., USA, 3111-3119*, Lake Tahoe, Nevada, 2013.

[4] M. Grbovic, V. Radosavljevic, N. Djuric, N. Bhamidipati, J. Savla, V. Bhagwan and D. Sharp, "E-commerce in Your Inbox: Product Recommendations at Scale," in *proceedings of the 21th ACM SIGKDD International Conference on Knowledge Discovery and Data Mining (KDD '15). ACM, New York, NY, USA, 1809-1818. DOI: https://doi.org/10.1145/2783258.2788627* , Sydney, NSW, Australia, 2015.

[5] R. He and J. McAuley, "Fusing similarity models with markov chains for sparse sequential recommendation," in *Proceedings of ICDM'16,*, 2016.

[6] B. Hidasi, A. Karatzoglou, L. Baltrunas and D. Tikk, "Session based recommendations with recurrent neural networks," in *Proceedings of ICLR'16*, 2016.

[7] Y. Tan, X. Xu and Y. Liu, "Improved recurrent neural networks for session-based recommendations," in *Proceedings of the 1st Workshop on Deep Learning for Recommender Systems*, 2016.

[8] R. Devooght and H. Bersini, "Long and short-term recommendations with recurrent neural networks," in *Proceedings of the 25th Conference on User Modeling, Adaptation and Personalization*, 2017.

[9] I. Sutskever, V. Oriol and V. L. Quoc, "Sequence to sequence learning with neural networks," *Advances in neural information processing systems,* pp. 3104-3112, 2014.

[10] U. TANIELIAN, M. GARTRELL and F. VASILE, "Adversarial Training of Word2Vec for Basket

Completion," *arXiv preprint arXiv:1805.08720, 2018,* 2018.

[11] R. DEVOOGHT and H. BERSINI, "Accelerating model-based collaborative filtering with item clustering," in *2018 International Joint Conference on Neural Networks (IJCNN)*, 2018.

[12] T. Bai, P. Du, W. X. Zhao, J. R. Wen and J. Y. Nie, "A Long-Short Demands-Aware Model for Next-Item Recommendation," arXiv preprint arXiv:1903.00066, Montreal, 2019.

[13] A. Sordoni, Y. Bengio, H. Vahabi, C. Lioma, J. Grue Simonsen and J. Y. Nie, "A Hierarchical Recurrent Encoder-Decoderfor Generative Context-Aware Query Suggestion," in *Proceedings of the 24th ACM International on Conference on Information and Knowledge Management.*, 2015.

[14] J. Chung, Ç. Gülçehre, K. Cho and Y. Bengio, "Empirical Evaluation of Gated Recurrent Neural Networks on Sequence Modeling," *CoRR,* vol. abs/1412.3555, 2014.

[15] D. Rumelhart, G. E. Hinton and R. J. Williams, "Learning representations by back-propagating errors," *Nature,* vol. 323, pp. 533-536, 1986.

[16] P. Webros, "Backpropagation through time: what it does and how todo it," in *Proceedings of IEEE*, 1990.

[17] L. Piciu, A. Damian, N. Tapus, A. Simion-Constantinescu and I. B. Dumitrescu, "Deep recommender engine based on efficient product embeddings neural pipeline," *2018 17th RoEduNet Conference: Networking in Education and Research (RoEduNet),* pp. 1-6, 2018.

[18] A. Simion-Constantinescu, I. A. Damian, N. Tapus, L.-G. Piciu, A. Purdila and B. Dumitrescu, "Deep Neural Pipeline for Churn Prediction," *10.1109/ROEDUNET.2018.8514153,* pp. 1-7, 2018.

[19] S. Hochreiter and J. Schmidhuber, "Long short-term memory," *Neural Comput. 9,* p. 1735–1780, 1997.

[20] M. Abadi, A. Agarwal, P. Barham, E. Brevdo, Z. Chen, C. Citro, G. Corrado, A. Davis, J. Dean, M. Devin, S. Ghemawat, I. Goodfellow, A. Harp, G. Irving, M. Isard, Y. Jia, L. Kaiser, M. Kudlur, J. Levenberg and X. Zheng, "TensorFlow : Large-Scale Machine Learning on Heterogeneous Distributed Systems," 2015.

[21] S. Wang, L. Hu, L. Cao, X. Huang, D. Lian and W. Liu, "Attention-based transactional context embedding for next-item recommendation," in *Thirty-Second AAAI Conference on Artificial Intelligence*, 2018.

[22] T. Bai, J.-Y. Nie, W. X. Zhao, Y. Zhu, P. Du and J.-R. Wen, "An attribute-aware neural attentive model for next basket recommendation," in *The 41st International ACM SIGIR Conference on Research \& Development in Information Retrieval*, ACM, 2018, pp. 1201--1204.

[23] A. Vaswani, N. Shazeer, N. Parmar, J. Uszkoreit, L. Jones, A. N. Gomez, L. Kaiser and I. Polosukhin, "Attention is All you Need," in *Advances in Neural Information Processing Systems 30*, Curran Associates, Inc., 2017, pp. 5998-6008.